\documentclass[12pt]{iopart}
\pdfoutput=1 
\usepackage[T1]{fontenc} 
\usepackage{mathrsfs, amssymb,amsfonts,amsthm,graphicx, epsf, yfonts, array}
\expandafter\let\csname equation*\endcsname\relax
\expandafter\let\csname endequation*\endcsname\relax
\usepackage{amsmath}
\usepackage{xcolor}
\usepackage{enumitem} 
\usepackage{cite}
\usepackage{changepage} 
\usepackage[hyperfootnotes=true,hyperindex,breaklinks]{hyperref}
\usepackage{color}
\usepackage{orcidlink}
\usepackage[utf8]{inputenc}
\DeclareMathAlphabet{\mathpzc}{OT1}{pzc}{m}{it}
\newcommand{\textoperatorname}[1]{%
  \operatorname{\textnormal{#1}}%
}
\usepackage{geometry}
\def\Z#1{_{\lower2pt\hbox{$\scriptstyle#1$}}}
\def\goesas{\mathop{\sim}\limits}
\def\lsim{\mathop{\hbox{${\lower3.8pt\hbox{$<$}}\atop{\raise0.2pt\hbox{$\sim$}}$}}}

\def\vD{v\Z{D}} \def\vK{v\Z{K}} \def\vd{v\Z{d}} \def\vh{v\Z{h}}  \def\vb{v\Z{b}} \def\vB{v\Z{B}}
\def\vN{v\Z{N}}
\def\vKzz{v\Z{K,zz}}
\def\vhzz{v\Z{h,zz}}
\def\vhr{v\Z{h,r}}
\def\PN{\Phi\Z{N}} \def\PD{\Phi\Z{D}}
 
\def\PDr{\Phi\Z{D,r}} \def\PDz{\Phi\Z{D,z}}
\def\pzz{\tilde{p}\Z{,zz}}
\def\pz{\tilde{p}\Z{,z}}
\def\pr{\tilde{p}\Z{,r}}
\def\dd{\mathop{\text{d}\!}}

\def\Md{M\Z d} \def\rd{r\Z d} \def\zd{z\Z d} \def\RDM{R\Z {DM}} \def\Mb{M\Z b} \def\Rb{R\Z b}
\def\rhoB{\rho\Z B} \def\rhoDM{\rho\Z {DM}} \def\rhoM{\rho\Z M}  \def\rhob{\rho\Z b} 
\def\rhod{\rho\Z d}
\def\rhodr{\rho\Z{d,r}} 
\def\rhodz{\rho\Z{d,z}}

\begin{document}
\title[Quasilocal Newtonian limit of general relativity and galactic dynamics]{Quasilocal Newtonian limit of general relativity\\ and galactic dynamics}

\author{Marco Galoppo\footnote{Author to whom any correspondence should be addressed.}$^1$\orcidlink{ 0000-0003-2783-3603}, Federico Re$^{2,3}$\orcidlink{0000-0002-9528-0855}, and David L.~Wiltshire$^1$\orcidlink{0000-0003-1992-6682}}

\address{$~1$ School of Physical \& Chemical Sciences, University of Canterbury, \\ Private Bag 4800, Christchurch 8140, New Zealand}
\address{$^2$ Dipartimento di Fisica ``Giuseppe Occhialini'', Universit\`{a} di Milano Bicocca,\\ Piazza dell'Ateneo Nuovo 1, 20126, Milano, Italy}
\address{$^3$ INFN, sezione di Milano, Via Celoria 16, 20133, Milano, Italy}

\eads{\mailto{Marco.Galoppo@canterbury.ac.nz}, \mailto{Federico.Re@unimib.it},\\ and \mailto{David.Wiltshire@canterbury.ac.nz}}

\begin{abstract}
\begin{adjustwidth}{0cm}{4cm}
{We present a new self-consistent perturbative expansion for realistic isolated differentially rotating systems -- disc galaxies. At leading order it is formally equivalent to Ehlers' Newton-Cartan limit, which we reinterpret in terms of quasilocal energy and angular momentum. The self-consistent coupling of these quasilocal terms leads to first-order differences from the conventional Newtonian limit. A modified Poisson equation is obtained, along with modifications to the equations of motion for the effective fluid elements. By fitting to astrophysical data, we show that the phenomenology of collisionless dark matter for disc galaxies can be reproduced. Potential important consequences for gravitational physics on galactic and cosmological scales are briefly discussed.}
\end{adjustwidth}
\end{abstract}
\CQG{\bf42} (2025) 135004\\ {\bf DOI}: 10.1088/1361-6382/ade48b\bigskip
\maketitle

\section{Introduction}

The Newtonian limit of general relativity is crucially important to both fundamental physics and astrophysical phenomenology. In cosmology two key features consistent with observation are:
\begin{enumerate}[label=(\roman*)]
    \item far from any localised sources the Universe is not empty; \item  the Universe is not globally rotating on the largest scales.
\end{enumerate}
Conventionally, one invokes an asymptotic Minkowski space, with respect to which additive Newtonian potentials are defined, {in spite of} observation (i). Moreover, even though known rotating asymptotically flat solutions of Einstein's equations possess global non-zero angular momenta at spatial infinity, it is implicitly assumed that superposition of the effectively quasilocal angular momenta of such pointlike sources can be reconciled with observation (ii).

In this paper we show that such implicit assumptions {overlook essential features of general relativity,} even at nonrelativistic orbital speeds. The conventional limit neglects an important coupling of the quasilocal energy and angular momentum defined by the regional time--averaged motion of matter sources. {For rotating galaxies we} set out a new, self-consistent \emph{quasilocal Newtonian limit of general relativity}, which does not assume a Minkowskian spacetime background \textit{a priori}. It introduces novel, first-order features that fundamentally modify {galactic dynamics}.

{The quasilocal Newtonian limit that we present coincides formally with the Newton-Cartan (NC) limit of general relativity introduced by J. Ehlers \cite{Ehlers_1981}. However, the previous physically relevant examples given by Ehlers -- including the Schwarzschild, Kerr and Friedmann-Lema\^{\i}tre-Robertson-Walker spacetimes \cite{Ehlers_1997}  -- 
return vanishing first-order quasilocal corrections for what Ehlers calls the ``Coriolis field''. Here, we apply Ehlers' Newton-Cartan limit to a realistic differentially rotating galaxy, with a fit to actual data. This is the first ever physically relevant nontrivial example of the Newton--Cartan limit of the Einstein equations. The explicit correspondence between our formalism and the original dictionary employed by Ehlers is given elsewhere \cite{Re_2025}.}

The prescription of a fixed, global spacetime background is not a requirement of general relativity. Indeed, the nonlinearity of Einstein's field equations (EFE), which arises from the self--interaction of matter and geometry, results in dynamical, regional backgrounds in which spacetime itself carries its own energy and angular momentum \cite{Szabados_2009,Wiltshire_2007,Wiltshire_2008,Wiltshire_2011}. Moreover, the highly nonlinear nature of general relativity gives rise to a noncommutativity of averaging and limiting procedures in spacetime \cite{Buchert_2000,Buchert_2020}. This has fundamental consequences for the dynamics of non-pointlike sources of large spatial extent, such as disc galaxies \cite{Astesiano_2022a,Astesiano_2022b,Re_2024,Galoppo_2024}.

The crucial ingredient of the differentially rotating models \cite{Astesiano_2022a,Astesiano_2022b,Astesiano_2022d,Astesiano_2022e, Ruggiero_2024, Re_2024,Galoppo_2024}, firstly considered in Ref.\ \cite{Astesiano_2022a}, and in particular in the exact solutions of Ref.\ \cite{Galoppo_2024}, is that essential nonlinearity is retained before any low energy limit is considered. No global background is assumed, {even in the linearised limit \cite{Astesiano_2022d,Astesiano_2022e}}. This contrasts with: {\bf(a)} approaches that investigate frame--dragging on asymptotically flat \cite{Ciotti_2022} or asymptotically Friedmann \cite{Bruni_2013,Thomas_2015} backgrounds; {\bf(b)} phenomenological models which modify the laws of gravitation, e.g., modified Newtonian dynamics (MOND) \cite{Milgrom_1983,Famaey_2012}.

In this paper we demonstrate how insights from the conventional Newtonian framework may nonetheless be embedded in a quasilocal framework with an effective fluid. Furthermore, Einstein was never fully satisfied by the extent to which the EFE embody Mach's principle, viz., \emph{``Local inertial frames are determined through the distributions of energy and momentum in the universe by some weighted average of the apparent motions''} \cite{Bondi_1961,Bicak_2007}. The quasilocal Newtonian limit exhibits such a suitable weighted average, which we apply to disc galaxies showing that the EFE may actually contain an understanding of inertia {consistent with the idea that MOND phenomenology may involve revisiting concepts about inertia \cite{Milgrom_2022}.}

\section{General relativistic galaxy metric}
The spacetime metrics can be written in the generalized Lewis--Papapetrou--Weyl form \cite{Re_2024,Galoppo_2024,Stephani_2003},
\begin{align}
    \label{eq:metric}
    \dd s^2 & =-c^2 e^{2\Phi(r, z)/c^2}(\dd t+A(r, z)  \dd\phi)^2 +e^{-2\Phi(r, z)/c^2}\left[W(r,z)^2 \dd\phi^2 +e^{2k(r, z)/c^2}(\dd r^2+\dd z^2)\right]\, .
\end{align}
The energy-momentum tensor takes the form
\begin{equation}
\label{eq:T}
    T^{\mu\nu}=\left(\rhoM(r, z)+p(r,z)/c^2\right) \, U^{\mu}U^{\nu}\, + p(r,z) g^{\mu\nu},
\end{equation}
where $\rhoM(r,z)$ is the local matter density, $p(r,z)$ is the effective pressure, and each element of the fluid possesses a 4--velocity $U^\mu$, given by
\begin{equation}
\label{eq:U}
    U^{\mu}\partial_\mu=(-H(r,z))^{-1/2}\left(\partial_t +\Omega(r,z) \, \partial_{\phi}\right)\, .
\end{equation}
Here $\Omega(r, z):= \dd \phi/ \dd t$ uniquely defines the angular speed of rotation at any point, {when pulled-back to the worldline of the fluid source,} and $H(r, z)$ is a normalization factor. Since $U^{\mu}U_{\mu}=-c^2$, it follows that
\begin{equation}
\label{eq:dustnorm}
     H=-e^{2\Phi/c^2}(1+A \, \Omega)^2+e^{-2\Phi/c^2}W^2  \, \Omega^2/c^2\, .
\end{equation}

Let us then write the EFE as
\begin{equation}
\label{eq:EinsteinsEquations}
    R_{\mu\nu} = \frac{8\,\pi\,G}{c^4}\left(T_{\mu\nu}-\frac{1}{2}T\,g_{\mu\nu}\right)\, ,
\end{equation}
where $T={T^\mu}_\mu = 3p-\rhoM \,c^2$. With \eqref{eq:metric}--\eqref{eq:dustnorm} this yields the partial differential equations (PDEs) \cite{Stephani_2003}
\allowdisplaybreaks
\begingroup
\setlength{\jot}{10pt} 
\setlength{\arraycolsep}{2pt}
\begin{align}
    & {\Phi^{,a}}_{\!,a} + \frac{W^{,a}\Phi_{,a}}{W} + c^4\frac{A^{,a}A_{,a}}{2W^2}e^{4\Phi/c^2} = 4\,\pi\,G e^{2(k-2\Phi)/c^2}\nonumber\\ &\hbox to 5cm{\hfil}\times\left[\left(\rhoM+\frac{p}{c^2}\right)\frac{(1+A\Omega)^2e^{2\Phi/c^2}+c^{-2}W^2\Omega^2e^{-2\Phi/c^2}}{-H}+2\frac{p}{c^2}\right] , \label{eq:tt_exact}\\
    &{A^{,a}}_{\!,a} - \frac{W^{,a}A_{,a}}{W} + \frac{4}{c^2}\Phi^{,a}A_{,a} = \frac{16\,\pi\,G}{c^4} W^2\Omega\frac{1+A\Omega}{H}
   \left(\rhoM+\frac{p}{c^2}\right)e^{2(k-2\Phi)/c^2} \, , \label{eq:tphi_exact}\\
   & {W^{,a}}_{\!,a} = \frac{16\,\pi\,G}{c^4}p \, , \label{eq:phiphi_exact}\\
   & W_{,rr}-W_{,zz}+\frac{2}{c^2}(k_{,z}W_{,z}-k_{,r}W_{,r})+\frac{2W}{c^4}\left(\Phi_{,r}^2 - \Phi_{,z}^2\right) + \frac{c^2}{2W}e^{4\Phi/c^2}\left(A_{,z}^2 - A_{,r}^2\right) = 0 \, , \label{eq:rrzz_exact}\\
   & \frac{c^2}{2W}e^{4\Phi/c^2}A_{,z}\,A_{,r}-W_{,rz}-\frac{2W}{c^4}\Phi_{,r}\Phi_{,z}+\frac{1}{c^2}(k_{,z}W_{,r}-k_{,r}W_{,z}) = 0 \, , \label{eq:rz_exact}\\
   & \frac{1}{c^2}\left({\Phi^{,a}}_{\!,a}-{k^{,a}}_{\!,a}\right)-\frac{1}{2W}{W^{,a}}_{\!,a} + \frac{1}{c^2}\frac{W^{,a}\Phi_{,a}}{W} - \frac{1}{c^4}\Phi^{,a}\Phi_{,a}+ \frac{c^2}{4W^2}e^{4\Phi/c^2}\left(A_{,z}^2 + A_{,r}^2\right)\nonumber\\ &\hbox to95mm{\hfil}= \frac{4\,\pi\,G}{c^2}e^{2(k-2\Phi)/c^2}\left(\rhoM-\frac{p}{c^2}\right) \, \label{eq:zzrr_exact},
\end{align} 
\endgroup
where $a\in\{r,z\}$. We note that \eqref{eq:tt_exact} and \eqref{eq:rz_exact} are the $tt$ and $rz$ components of \eqref{eq:EinsteinsEquations} respectively, \eqref{eq:rrzz_exact} and \eqref{eq:zzrr_exact} are obtained from the $R_{zz}\pm R_{rr}$ equations, \eqref{eq:phiphi_exact} is found by taking the combination $W^{-1}c^{-2}\left(g_{\phi\phi}R_{tt}-2\,g_{t\phi}R_{t\phi}+g_{tt}R_{\phi\phi}\right)$, and finally \eqref{eq:tphi_exact} follows from the combination $2c^{-2}\left(R_{t\phi}-A\,R_{tt}\right)$. 

The system of PDEs is completed by the perfect fluid elements' equations of motion, which give 
\begin{align}
        H\frac{p_{,a}}{\rhoM}e^{2(k-\Phi)/c^2}=\bigl[\Phi_{,a}+(c^2 A_{,a}+2A\Phi_{,a})\Omega+(c^2AA_{,a}&+A^2\Phi_{,a})\Omega^2\bigr]e^{2\Phi/c^2}\nonumber\\ &+\Omega^2\left(W^2\frac{\Phi_{,a}}{c^2}-WW_{,a}\right)e^{-2\Phi/c^2} \, .\label{eq:EoM}
\end{align}
\section{The quasilocal Newtonian limit}
Relevant physical velocities have to be identified to correctly implement any low-velocity limit. In the present case, we define the kinetic and dragging velocities
\begin{align}
    & \vK := r\,\Omega \, , \label{eq:vK}\\ 
    & \vD := r\,\chi \, , \label{eq:vD}
\end{align}
where $\chi := -g_{t\phi}/g_{\phi\phi}$ is the frame--dragging term. For systems in the low-energy regime---i.e., nonrelativistic relative local velocities $v \ll c$, weak pseudo-Newtonian potential $\Phi \goesas v^2$, nonrelativistic frame--dragging, and small pressure $p \goesas \rhoM v^2$---then to leading order $\vK$ coincides with the special relativistic interpretation of the redshift, whilst $\vD$ follows by analogy \cite{Astesiano_2022b,Re_2024,Galoppo_2024}.

We now take an expansion in powers of $v/c$ of Eqs. \eqref{eq:tt_exact}-\eqref{eq:EoM} to implement the nonrelativistic limit, where $v$ is any relevant local velocity\footnote{{In the low-energy limit of the class of spacetimes we consider, it necessarily follows that $\vD\goesas \vK \goesas v \ll c$. See Ref.\ \cite{Re_2024} for a more in-depth discussion.}}, {generalising the approach of \cite{Astesiano_2022d,Astesiano_2022e} to now include pressure}. We apply the self-consistent ansatz 
\begin{equation}\label{eq:WAnsatz}
    W(r,z) = r + \mathcal{O}\left(v^4/c^4\right) \, ,
\end{equation}
to solve \eqref{eq:phiphi_exact} up to fourth order. From \eqref{eq:metric}, and the definition of $\chi$, we find $A = \left(r\,\vD /c^2\right)\left[1+\mathcal{O}({v^3/c^3})\right]$. Thus, by \eqref{eq:dustnorm} it follows that 
\begin{equation} \label{eq:HLowVel}
    H = -1 + \frac{\vK^2}{c^2} - 2\frac{\vK}{c}\frac{\vD}{c} + \frac{2\,\Phi}{c^2} + \mathcal{O}\left(v^4/c^4\right)\,.
\end{equation}
Hence, by direct substitution of \eqref{eq:WAnsatz} into \eqref{eq:tt_exact}--\eqref{eq:EoM} we find
\begin{align}
    &\Delta \Phi + \frac{1}{2r^2}||\vec{\nabla}{\textoperatorname{Ł}_D}||^2 = 4\,\pi\,G \rhoM + \mathcal{O}\left(v^2/c^2\right) \,, \label{eq:ttexp}\\
    & \hat{\Delta}\textoperatorname{Ł}_D = 0 + \mathcal{O}\left(v^2/c^2\right) \, , \label{eq:tphiexp}\\
    & k_{,r} = \frac{1}{4r}\left(\textoperatorname{Ł}_{D,z}^2-\textoperatorname{Ł}_{D,r}^2\right)+ \mathcal{O}\left(v^2/c^2\right)  \, , \label{eq:rrexp}\\
    & k_{,z} = -\frac{1}{2r}\left(\textoperatorname{Ł}_{D,z}\,\textoperatorname{Ł}_{D,r}\right) + \mathcal{O}\left(v^2/c^2\right) \, , \label{eq:zzexp}
\end{align}
where $\textoperatorname{Ł}_D := r\,\vD$ is the quasilocal angular momentum per unit of mass associated with spacetime rotation, $\Delta$ is the standard Laplacian and $\hat{\Delta} := \partial^2_r -(1/r)\partial_r + \partial^2_z$, is the Grad--Shafranov Laplacian \cite{Ruggiero_2024,Re_2024,Galoppo_2024}. The number of equations reduces by one, as \eqref{eq:tt_exact} is equivalent to \eqref{eq:zzrr_exact} to the order considered. Furthermore, we see that the fluid effective pressure does not enter the EFE as a source term at this order of approximation. 

Equation \eqref{eq:ttexp} is a general relativistic generalization of the Poisson equation, and coincides with it in the absence of spacetime rotation. The extra term in \eqref{eq:ttexp} can be interpreted as the rotational energy associated with the quasilocal angular momentum of the averaged background spacetime. This term is absent in the conventional Newtonian approximation in which the background is taken to be Minkowski space \textit{a priori}. Furthermore, when moved to the r.h.s.\ of \eqref{eq:ttexp}, it can be identified as an effective density, $\rho\Z{\mathcal{Q}}=-||\vec{\nabla}{\textoperatorname{Ł}_D}||^2/(8\pi G\, r^2)$.  This represents a negative gravitational binding energy relative to the background \cite{Galoppo_2024}.

Even for nonrelativistic dragging velocities spacetime rotation gives first-order corrections to the field equations. For fixed $\rhoM$, \eqref{eq:ttexp} shows that the strength of the pseudo-Newtonian potential needed to sustain equilibrium decreases as the local rotational energy of the background geometry increases. Furthermore, the pseudo-Newtonian potential in \eqref{eq:ttexp} is essential to accommodate nonrigid rotation of the perfect fluid source. Indeed, if we set $\Phi=0$ then \eqref{eq:ttexp}--\eqref{eq:zzexp} are the same equations that characterize a system of rigidly rotating dust \cite{Stephani_2003,Balasin_2008,Crosta_2020,Beordo_2024}. {Although rigidly rotating models were applied in the first attempts to find nontrivial solutions of the EFE representing disc galaxies \cite{Balasin_2008,Crosta_2020,Beordo_2024}, it was always recognised that they were at best toy models as real disc galaxies are observed to be differentially rotating. Furthermore, when rigidly rotating GR models are applied to the dynamics of distant disc galaxies, then one is led to the paradoxical conclusion that the rotation velocity inferred from the redshift would be zero. (See e.g., \cite{Astesiano_2022b,Re_2024}.)}

{Expanding the equations of motion in the same procedure we applied to the EFE, at the same order in the expansion parameter we find that
 \eqref{eq:EoM} reduces} to 
\begin{align}
        &-\frac{p_{,r}}{\rhoM}=\Phi_{,r}+\Omega\,\textoperatorname{Ł}_{D,r} - \vK^2/r + \mathcal{O}\left(v^2/c^2\right) \, , \label{eq:EoMrapp}\\
        &-\frac{p_{,z}}{\rhoM}=\Phi_{,z}+\Omega \textoperatorname{Ł}_{D,z}+\mathcal{O}\left(v^2/c^2\right) \, \label{eq:EoMzapp}.
\end{align}
Hence, the effective pressure of the fluid still plays a crucial role in determining the physics of the system, as expected. {The pressure terms model the velocity dispersion within the disc galaxy, supporting the thickness of the galactic disc, and are crucial in determining its stability. In standard Newtonian analyses, the local stability of a disc galaxy is conventionally assessed via the Toomre $Q(r)$ stability parameter \cite{Toomre_1964,Binney_2008,Bertin_2014}. A value $Q(r) > 1$ indicates that a matter ring at radius $r$ is stable against axisymmetric perturbations of all wavelengths, whilst $Q(r) \gg 1$ throughout the galaxy indicates global stability to all linear perturbations \cite{Binney_2008,Bertin_2014}. Interestingly, the Toomre $Q$ parameter is automatically zero for systems that are either rigidly rotating or have no velocity dispersion. Since the terms responsible for driving the positivity of the Toomre parameter -- differential rotation and internal effective pressure -- are both present in the quasilocal Newtonian limit, we expect that general relativistic disc galaxy models can be found which are locally, and possibly globally, dynamically stable. Furthermore, the introduction of an effective pressure allows us to remove any ambiguities in the choice of galactic density profiles which are present even in differentially rotating, pressureless GR galaxy models \cite{Re_2024}.}

In \eqref{eq:EoMrapp} and \eqref{eq:EoMzapp}, we recognize the gradient of the pseudo-Newtonian potential and the centrifugal acceleration in the term $\vK^2/r$. The conventional Newtonian equations follow in the case of negligible dragging ($\vD\ll\vK$). However, more generally \eqref{eq:EoMrapp} and \eqref{eq:EoMzapp} once again show the crucial role that the quasilocal angular momentum of the regional background plays in determining the dynamics. For fixed $\rhoM$ and $p$, the magnitude of the pseudo-Newtonian potential needed to match the l.h.s.\ of \eqref{eq:EoMrapp} and \eqref{eq:EoMzapp} decreases linearly with the quasilocal angular momentum of the underlying averaged geometry. {Thus \eqref{eq:ttexp}-\eqref{eq:EoMzapp} define the quasilocal Newtonian limit.}

\section{Application to a galactic model}

As a first application of the quasilocal Newtonian limit, we derive the dragging velocity profile for a disc galaxy rotation curve supported exclusively by baryonic matter. The baryonic mass distribution in a disc galaxy can be written as $\rhoB = \rhob + \rhod$, where $\rhob$ and $\rhod$ are the densities of the bulge and the disc, respectively. 

We assume a spherical bulge component given by the Plummer density profile \cite{Pouliasis_2017}
\begin{equation}
    \rhob (R) = \frac{3\Rb^2\Mb}{4\pi\left(R^2+\Rb^2\right)^{5/2}} \,,
\end{equation}
where $R = \sqrt{r^2+z^2}$, $\Rb$ is the scale parameter of the bulge and $\Mb$ is the total bulge mass. The resulting circular velocity in the conventional Newtonian limit is
\begin{equation}
    \vb(R) = \sqrt{G\Mb R^2\left(R^2+\Rb^2\right)^{-3/2}}\, .
\end{equation}

The baryonic disc matter distribution is usually taken as $\rhod(r, z)=\Sigma(r)Z(z)$. The surface density of the disc is well approximated by the exponential profile
\begin{equation}
\label{eq:rdensity}
    \Sigma(r)=\frac{\Md}{2\pi \rd^2}e^{-r/\rd}\, ,
\end{equation}
where $\Md$ is the total disc mass and $\rd$ is the scale length in the radial direction for the galactic disc. The distribution along $z$, with $\int Z(z)dz=1$, is assumed to be
\begin{equation}
\label{exp z}
    Z(z)=\frac{1}{2\zd}e^{-|z|/\zd} \, ,
\end{equation}
where $\zd$ is the thickness scale length and $\zd \ll \rd\,$. When the thickness $\zd$ tends to zero, we apply the conventional thin disc approximation, i.e., $Z(z)=\delta(z)$\cite{Ciotti_2022,Freeman_1970}. The Newtonian circular velocity field in the thin disc approximation, $\vd$, is given by \cite{Ciotti_2022,Freeman_1970}
\begin{equation}
    \label{eq:vd}
    \vd(r) = \sqrt{\frac{G\Md}{2\rd^3}r^2\left[I_0\left(\frac{r}{2\rd}\right)K_0\left(\frac{r}{2\rd}\right) -I_1\left(\frac{r}{2\rd}\right)K_1\left(\frac{r}{2\rd}\right)\right]}\, ,
\end{equation}
where $I_{\{0,1\}}$ and $K_{\{0,1\}}$ are modified Bessel functions of the first and second kind.

\begin{figure}[htb!]
    \centering
    \includegraphics[]{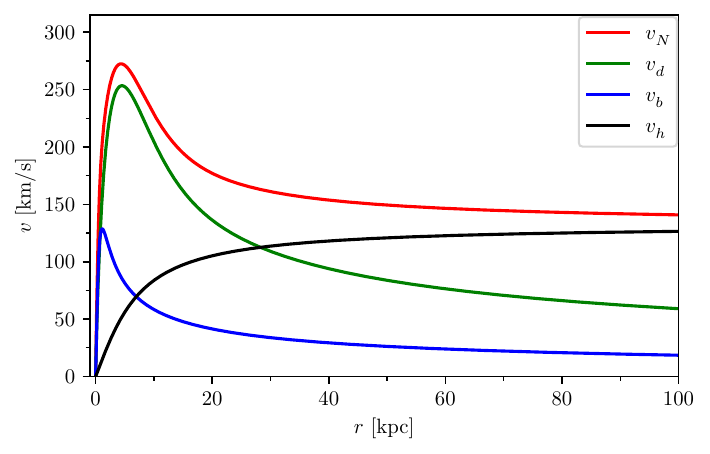}
    \caption{Conventional Newtonian rotation curve of the thin disc, $\vK(r,0)$, and the respective $\vb(r,0)$, $\vd(r)$ and $\vh(r,0)$, for a Milky Way-like galaxy with  $\Mb = 0.8 \cdot 10^{10}\,$M$_\odot$, $\Rb = 0.8$ kpc, $\Md = 8.1 \cdot 10^{10}\,$M$_\odot$, $\rd = 2.1$ kpc, $\RDM = 5.69$ kpc and $\rho\Z{DM0}= 6.77\cdot10^{-22}$ kg/m$^3$.}
    \label{fig:figureN}
\end{figure}

To explain the observed rotation curves in the conventional Newtonian framework, a spherical cold dark matter (DM) halo must be included, so that the matter density is $\rhoM = \rhoB + \rhoDM$, where $\rhoDM$ is typically given by the isothermal profile \cite{Kent_1986,Begeman_1991,Re_2024}
\begin{equation}
    \label{eq:rhoDM}
    \rhoDM(R)=\frac{\rho\Z{DM0}}{1+(R/\RDM)^2}\, .
\end{equation}
Here $\RDM$ is the halo scale parameter, and $\rho\Z{DM0}$ is the maximum dark matter density. The dark matter halo contribution to the rotation curve, $\vh$, is then \cite{Re_2024}
\begin{equation}
    \vh(R)^2=4\pi G\, \RDM^2\,\rho\Z{DM0} \left(1- \frac{\arctan{\left(R/\RDM
\right)}}{R/\RDM}\right) \, .
\end{equation}
Thus the circular velocity of the matter in Newtonian modelling of the disc galaxy, $\vN$, on the equatorial plane, is given by
\begin{equation}
    \vN(r,0)=\sqrt{\vb(r,0)^2+\vd(r)^2+\vh(r,0)^2}\, .
\end{equation}
In figure \ref{fig:figureN}, we consider a Milky Way-like galaxy with $\Mb = 0.8 \cdot 10^{10}\,$M$_\odot$, $\Rb = 0.8$ kpc, $\Md = 8.1 \cdot 10^{10}\,$M$_\odot$, $\rd = 2.1$ kpc, $\RDM = 5.69$ kpc and $\rho\Z{DM0}= 6.77\cdot10^{-22}$ kg/m$^3$. We plot the predicted rotation curve $\vK(r,0)$, and the respective $\vb(r,0)$, $\vd(r)$ and $\vh(r,0)$. 

To apply the general relativistic solution, let us now assume that $\vK = \vN$, i.e., that the inferred Newtonian rotation curve corresponds to that of a disc galaxy, and that $\rhoM = \rhoB$. We will demonstrate that the quasilocal energy and angular momentum of the galaxy produce observations equivalent to those of a conventional cold dark matter halo.

We begin by writing the pseudo-Newtonian potential as $\Phi=\PN-\PD$, so that $\PN$ is given by the Newtonian solution for the chosen $\rhoB$, while $\PD$ is defined as the solution of the Poisson equation sourced purely by the quasilocal rotational energy of the spacetime 
\begin{equation}
\label{eq:PD}
\Delta\PD=\frac{||\vec{\nabla}\textoperatorname{Ł}_D||^2}{2r^2}\, .
\end{equation}

In the thin disc approximation a detailed calculation shows that in the general relativistic case the corrections to the pressure term do not affect the dragging velocity of the system (see Appendix A), allowing a self-consistent solution of \eqref{eq:ttexp}, \eqref{eq:EoMrapp}, and \eqref{eq:EoMzapp} for $\textoperatorname{Ł}_D$.
Furthermore, a similar ansatz can be applied to the spherical bulge, where dark matter is not required to explain the observed dynamics, being well captured by conventional Newtonian models with purely baryonic matter. Therefore, to simplify the calculations, we can thus effectively take $p(r,z):=p\Z N (r,z)$ for the whole system, where $p\Z N (r,z)$ is the expected effective Newtonian pressure. We then use \eqref{eq:EoMrapp} and \eqref{eq:EoMzapp} to deduce
\begin{align}
        &\PDr = \Omega\textoperatorname{Ł}_{D,r} -\left(\vK^2-\vB^2\right)/r \, , \label{eq:PDr}\\
        &\PDz = \Omega\textoperatorname{Ł}_{D,z}\, \label{eq:PDz},
\end{align}
where $\vB:= \sqrt{\vb^2+\vd^2}\,$ is the Newtonian contribution to the observed rotation curve from the baryonic mass.

Substituting \eqref{eq:PDr}, \eqref{eq:PDz} in \eqref{eq:PD} it follows that
\begin{align}
\label{eq:LDPDE1}
    \Delta\PD =\Omega_{,r}\textoperatorname{Ł}_{D,r} +2\Omega\frac{\textoperatorname{Ł}_{D,r}}{r} +\Omega_{,z}\textoperatorname{Ł}_{D,z} -\frac{\left(\vK^2-\vB^2\right)_{,r}}{r} \, ,
\end{align}
where we have also used the quasilocal Newtonian limit of the $(t\phi)$ EFE \eqref{eq:tphiexp}. Eqs.\ \eqref{eq:PD}, \eqref{eq:LDPDE1} then give a first-order nonlinear PDE for $\textoperatorname{Ł}_D$
\begin{equation}
\label{eq:LDPDE2}
  \vec{\nabla}{\textoperatorname{Ł}_D} \cdot \left[\vec{\nabla}{\textoperatorname{Ł}_D} - 2\vec{\nabla}{\left(r\vK\right)}\right] + 2r\left(\vK^2-\vB^2\right)_{,r} = 0\, .
\end{equation}

We evaluate \eqref{eq:LDPDE2} on the galactic plane, $z=0$, where due to the plane symmetry we have $\textoperatorname{Ł}_{D,z}(r,0)=v\Z{K,z}(r,0) = 0$. Thus, the nonlinear PDE \eqref{eq:LDPDE2} reduces to an algebraic equation for $\textoperatorname{Ł}_{D,r}$ on the galactic plane, with solutions given by 
\begin{equation}
\label{eq:PracticalLDr}
     \textoperatorname{Ł}_{D,r}=(r\vK)_{,r} \pm\sqrt{\left[(r\vK)_{,r}\right]^2 - 2r\left(\vK^2-\vB^2\right)_{,r}}\, .
\end{equation}
We take the minus sign in \eqref{eq:PracticalLDr} since it is the only choice consistent with a vanishing dragging velocity, $\vD\to0$, in the limit $r\to\infty$, as expected for a disc galaxy. It then reduces to the Newtonian case for $\vK=\vB$. Furthermore, given the baryonic density distribution and the conventional Newtonian profile for $\vK(r,0)$, we can numerically solve \eqref{eq:PracticalLDr} for $\vD(r,0)$.
The resulting planar solution for $\partial_r\textoperatorname{Ł}_{D}$ \eqref{eq:PracticalLDr}, and its integral $\textoperatorname{Ł}_D$ then supply a boundary condition to completely fix $\textoperatorname{Ł}_D$ throughout the spacetime via \eqref{eq:LDPDE2}. The resulting $\textoperatorname{Ł}_D(r,z)$ field then automatically satisfies the remaining EFE that were applied in the derivation of \eqref{eq:LDPDE2}. Thus by solving for $\vD$ on the galactic plane, we fix its value over the whole spacetime. 

\begin{figure}[htb!]
    \centering
    \includegraphics[width=0.63\columnwidth]{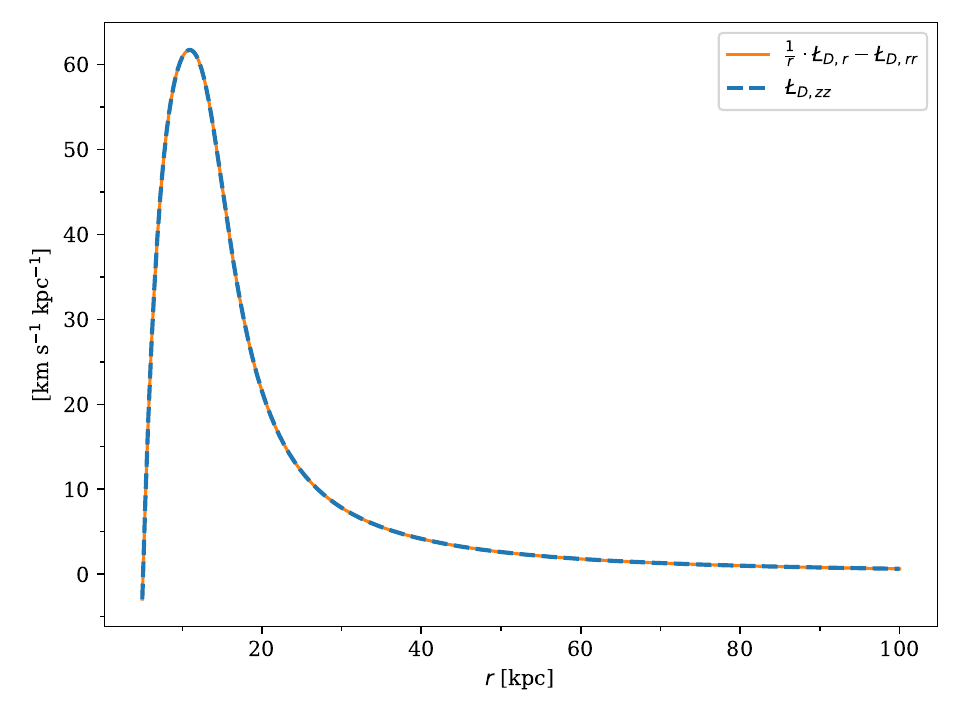}
    \caption{{The profiles for $\textoperatorname{Ł}_{D,r}/r - \textoperatorname{Ł}_{D,rr}$ (solid curve), and $\textoperatorname{Ł}_{D,zz}$ (dashed overlay) on the galactic plane,  for a Milky Way-like galaxy with $\Mb = 0.8 \cdot 10^{10}\,$M$_\odot$, $\Rb = 0.8$ kpc, $\Md = 8.1 \cdot 10^{10}\,$M$_\odot$ and $\rd = 2.1$ kpc.}}
    \label{fig:figureGS}
\end{figure}
{Figure \ref{fig:figureGS} shows the galactic plane profiles of: (i) $ \textoperatorname{Ł}_{D,r}/r - \textoperatorname{Ł}_{D,rr}$ determined from \eqref{eq:PracticalLDr}; (ii) $\textoperatorname{Ł}_{D,zz}$ determined from the systems of equations \eqref{eq:ttexp}, \eqref{eq:EoMrapp}, \eqref{eq:EoMzapp} which incorporate an approximation for the pressure. We see that the two profiles coincide up to numerical error, giving residuals of the order $10^{-14}$, so that $\textoperatorname{Ł}_{D}$ satisfies the Grad--Shafranov equation \eqref{eq:tphiexp} on the galactic plane. It will then also satisfy \eqref{eq:tphiexp} away from the galactic plane as \eqref{eq:tphiexp} is used in the derivation of the master equation \eqref{eq:LDPDE2}. In figure \ref{fig:figureGS} we start to observe divergences in both the profiles of $\textoperatorname{Ł}_{D,r}/r - \textoperatorname{Ł}_{D,rr}$ and $\textoperatorname{Ł}_{D,zz}$ as $r \rightarrow 0$. This divergence, which generates no physical effects, is simply related to the well-known divergence of the second radial derivative of the classical Newtonian potential at $r = 0$ for an exponential disc. The use of an alternative thin disc profile, such as a Kuzmin disc, would remove this divergence \cite{Binney_2008}.}

\begin{figure}[htb!]
    \centering
    \includegraphics[]{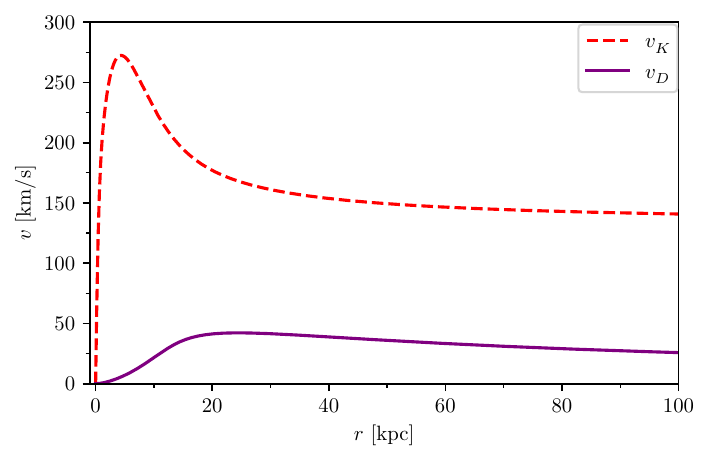}
    \caption{The profile for $\vD(r,0)$ needed to support the simulated rotation curve, $\vK(r,0)$,  for a Milky Way-like galaxy with $\Mb = 0.8 \cdot 10^{10}\,$M$_\odot$, $\Rb = 0.8$ kpc, $\Md = 8.1 \cdot 10^{10}\,$M$_\odot$ and $\rd = 2.1$ kpc. Here, the galaxy is composed exclusively of baryonic matter, i.e, $\rhoDM = 0$.}
    \label{fig:figureGR}
\end{figure}

Finally, in figure \ref{fig:figureGR} we show the frame--dragging velocity profile, $\vD$, which generates the same galaxy rotation curve of figure \ref{fig:figureN}, assuming $\vK=\vN$. The frame--dragging speed reaches a maximum $42.2\,$km/s at $r=24.4\,$kpc, well within our nonrelativistic approximation. Furthermore, as expected: {\bf(i)} beyond its maximum $\vD$ is monotonically decreasing with $r$;
and {\bf(ii)} $\vD$ is negligible in the bulge, showing the self-consistency of the pressure ansatz used in the calculations. {Here we have presented one explicit example of the phenomenological application of the quasilocal Newtonian limit of disc galaxies. However, \eqref{eq:LDPDE2} and \eqref{eq:PracticalLDr} may be consistently applied to all disc galaxies morphologies, e.g., also to bulgeless dwarf disc galaxies, thereby allowing for a direct consistency test of the phenomenology on actual astronomical data.}

\section{Conclusions}

{In this article we have established a self-consistent low-energy limit for an effective fluid model of astrophysically realistic isolated differentially rotating disc galaxies. Each fluid element contains many particles of stars and gas: what is important is not only their combined internal mass--energy and their non-gravitational interactions, but the quasilocal energy and angular momentum that arise from the gravitational interactions of the stars and gas within each effective fluid element. These determine their regional background geometry via the EFE \eqref{eq:tt_exact}--\eqref{eq:zzrr_exact} and the equations of motion \eqref{eq:EoM}. The low--energy limit we find is contained \cite{Re_2025} within Ehlers' Newton--Cartan limit \cite{Ehlers_1981,Ehlers_1997}.

Ehlers gave the name ``Coriolis field'' to the terms that arise from asymptotic frame--dragging. However, of the previously known spacetimes that Ehlers applied the Newton--Cartan limit to \cite{Ehlers_1997}, the only example with a nontrivial Coriolis field was the unphysical NUT spacetime. Ours is the first example of a physically significant Coriolis field, generated by differentially rotating shells of matter, with their nontrivial quasilocal angular momentum surviving in the low-energy limit. Most significantly, in this quasilocal Newtonian limit the standard Poisson equation is modified by an additional term according to \eqref{eq:ttexp}. }

{An interesting open question is how the additional quasilocal terms in \eqref{eq:ttexp}--\eqref{eq:zzexp} are to be interpreted in conventional post-Newtonian expansions, particularly in the parametrised post-Newtonian (PPN) formalism \cite{Will_2014,Poisson_2014}. The conventional Newtonian limit has been well confirmed within the standard PPN framework for few-body systems -- e.g., in the solar system and the dynamics of compact objects in binaries \cite{Will_2014,Poisson_2014}, as well as gravitational waves production in such systems \cite{Ligo_2021a,Ligo_2021b,Colleoni_2024}. On these scales any quasilocal corrections are expected to be negligible. Extensions of the PPN formalism to cosmological settings has produced interesting results with some degeneracies between the PPN and cosmological parameters \cite{Sanghai_2017, Thomas_2024}. Any cosmological PPN tests which a priori assume neither the dark matter content of a galactic halo nor the alternative quasilocal contributions, could potentially provide a direct means for distinguishing the two scenarios. We note that the $\Delta_{i = 1,2}$ and the $\beta_{i = 1,2}$ PPN parameters \cite{Will_2014,Poisson_2014} -- which are related to frame-dragging effects -- might be expected to be the PPN parameters most sensitive to quasilocal contributions on galactic scales.}

The new quasilocal Newtonian limit applies to stationary axisymmetric matter distributions with geometry \eqref{eq:metric}. These symmetry assumptions mean that for a Milky Way-like disc galaxy the approximations here, along with the exact solutions \cite{Galoppo_2024} hold on time scales, $10^7\lsim t\lsim10^9\,$yr. In particular, in this approximation we neglect the effective anisotropic quasilocal pressure from gravitational shear between the radial and angular directions. Volume-preserving deformations of the effective fluid elements driven by this term could play a role in generating galactic spiral arms and bars. However, these break the axial symmetry and stationarity of the models considered here, and require further development.

Likewise, equations \eqref{eq:ttexp}--\eqref{eq:EoMzapp} will not apply during galaxy formation, and therefore do not explain how disc galaxies form in a cold dark matter free environment. The purpose of all model building is to create the best superstructure upon which more complex problems can be tackled in future. In our view, we have established key elements of that superstructure by clearly identifying the relevant physical quantities. The precise definition of finite infinity \cite{Wiltshire_2007,Wiltshire_2008,Wiltshire_2011,Galoppo_2024} provides a mathematical framework for further exploring the quasilocal Newtonian limit, in settings from the early Universe through to highly dynamical astrophysical systems. The interplay of quasilocal gravitational energy and angular momentum must play a key crucial role.

In summary, we have shown that general relativity admits a new self-consistent low energy limit: the quasilocal Newtonian limit, {a nontrivial realization of Ehlers' Newton--Cartan limit \cite{Ehlers_1981,Ehlers_1997,Re_2025}}. In this limit the phenomenology of collisionless dark matter for disc galaxies can be reproduced by the regional gravitational energy and angular momentum of the average spacetime geometry.

This discovery {potentially} has far-reaching consequences for cosmology, astrophysics and particle physics. Indeed, many interesting questions are now opened up. {How do quasilocal energy and angular momentum contribute to the phenomenology of dark matter in the early universe, to the dynamics of disc galaxies, galaxy clusters, and other astrophysical environments?}

{Some environments involve a hierarchy of scales and related gravitational energy contributions to the fitting problem \cite{Wiltshire_2011}. The effective pressure and kinetic energy from the velocity dispersion of galaxies in galaxy clusters, with no large scale rotation, can be expected to be qualitatively and quantitatively different to the effective pressure and rotational energy and angular momentum of stars and gas in disc galaxies. Our aim in the present article has been to tackle the dynamics of stars and gas within disc galaxies. The more ambitious goal of addressing the next scale of fitting galaxies into galaxy clusters may potentially have a resolution in terms of quasilocal gravitational energy. However, much further development of the formalism is required before modelling systems such as the dynamics of the Bullet Cluster \cite{Douglas_2006,Freese_2017}.}

{The question of the stability of disc galaxies within the quasilocal Newtonian limit needs to be fully addressed.
Insofar as the velocity dispersion effective pressure for the general relativistic model of figure \ref{fig:figureGR} is identical to that producing figure \ref{fig:figureN} for Newtonian gravity, we can expect much of the standard Toomre $Q(r)$-parameter analysis to carry over to the quasilocal Newtonian limit. In the standard stability analyses, values $1.5 \lsim Q \lsim 2.5$ are found in the solar neighbourhood \cite{Binney_2008,Bertin_2014}, indicative of global stability. However, a full theoretical derivation of the Toomre stability parameter accounting for the quasilocal contributions is needed to check consistency.}

{In the past decade gravitational lensing data has been combined with dynamical data in massive disc galaxies to probe the properties of dark matter halos \cite{Trott_2010,Dutton_2011,Suyu_2012}, and to test modified gravity models such as MOND \cite{Mistele_2024a,Mistele_2024b}.
 Analogous tests could be performed to break the degeneracy between dark matter and quasilocal contributions to rotation curves. To do so, a formalism for gravitational lensing within the quasilocal Newtonian limit is required, generalizing that for rotating lenses in the post-Newtonian expansion \cite{Asada_2000,Sereno_2003a,Sereno_2003b}. Initial calculations have been performed \cite{Re_2024}, which have shown that the quasilocal corrections modify the bending angle by a similar order of magnitude as is inferred for cold dark matter.}

{Finally, dark matter density profiles are conventionally parameterised by a number of parameters that varies with the complexity of the profile, as an aid to observational data fitting. For the quasilocal contributions, an optimal functional parameterisation of either $\textoperatorname{Ł}_{D}$ or $\vD$ for different galactic morphologies needs to be developed. A starting point for this are investigations of fits to large samples of galaxy rotation curves.}
\newpage
\noindent{\bf Acknowledgments}\quad 
DLW and MG are supported by Marsden Fund grant M1271 administered by the Royal Society of New Zealand, Te Ap\=arangi. We are grateful to Davide Astesiano, Luca Ciotti, Tim Clifton, Sergio Cacciatori, John Forbes, Vittorio Gorini, Christopher Harvey-Hawes, Asta Heinesen, Frederic Hessman, Morag Hills, Emma Johnson, Zachary Lane, Pierre Mourier, Ryan Ridden-Harper, Antonia Seifert, Chris Stevens, Shreyas Tiruvaskar and Michael Williams for useful discussions. 
{MG and DLW thank participants at the Tsinghua Sanya International Mathematics Forum workshop, {\em Structures and Dynamics in Cosmology}, 13--17 January, 2025, for enlightening discussions -- and, in particular, J\"org Frauendiener for making us aware of Ehlers' discovery \cite{Ehlers_1981,Ehlers_1997} of the Newton--Cartan limit. FR thanks Oliver Piattella for joint work in relating Ehlers' formalism to ours \cite{Re_2025} and independently bringing \cite{Ehlers_1981} to our attention.} 
\\

\bibliographystyle{iopart-num}
\bibliography{Pressure.bib}

\appendix
\section{}
\label{A}
\renewcommand{\theequation}{A.\arabic{equation}}
\setcounter{equation}{0} 

Here we show that for a thin disc the general relativistic corrections to the pressure do not affect the frame-dragging velocity, i.e., Eqn. \eqref{eq:PracticalLDr} is independent of them. We start by considering $p=p\Z{N}+\tilde{p}$, where $\tilde{p}$ is the pressure  difference between the purely Newtonian and the full GR cases. Then \eqref{eq:EoMrapp} and \eqref{eq:EoMzapp} are found to take the form
\begin{align}
    &\PDr=\Omega\textoperatorname{Ł}_{D,r} -\frac{\vh^2}{r}  -\frac{\pr}{\rhod}\, , \\
    &\PDz=\Omega\textoperatorname{Ł}_{D,z} -\frac{\pz}{\rhod}\, .
\end{align}
Therefore, we have from \eqref{eq:tphiexp} and \eqref{eq:PD}  
\begin{align} 
\label{eq:A1}
    \Delta\PD = &\frac{\textoperatorname{Ł}_{D,r}^2+\textoperatorname{Ł}_{D,z}^2}{2r^2}= \Omega_{,r}\textoperatorname{Ł}_{D,r} +2\Omega\frac{\textoperatorname{Ł}_{D,r}}{r} +\Omega_{,z}\textoperatorname{Ł}_{D,z} \nonumber \\
    & -\frac{\Delta\tilde{p}}{\rhod}-\frac{(\vh^2)_{,r}}{r} + \frac{\pr\rhodr+\pz\rhodz}{\rhod^2} \, ,
\end{align}
The thin disc is obtained for vanishing pressure, i.e. when both $p\Z{N}(r, z)$ and $\tilde{p}(r, z)$ tends to zero in $L^1(\mathbb{R}^3)$. Indeed, in the Newtonian case it is known that for $\rho_d(r,z)=[\Sigma(r)//2\pi z_d]\eta(r,z/z_d)$, $p\Z{N}$ does not tend to zero pointwise, but is rather defined as \cite{Binney_2008}
\begin{equation}
\label{eq:A3}
    p\Z{N}(r, z)=\frac{\pi}{2}\,G\,\Sigma(r)^2\,\eta( r,z/\zd) \,,   
\end{equation}
where
\begin{equation}
\label{eq:A4}
    \eta(r,0)\equiv 1,\quad  \text{and} \quad \int_{-\infty}^{+\infty}\eta(r, s)ds<\infty\, ,
\end{equation}
Analogously, we will assume that the GR correction on the pressure tends to zero as
\begin{equation}
\label{eq:A5}
    \tilde{p}(r, z):=\tilde{p}_0(r)\zd^{\alpha}\tilde{\eta}(r,z/\zd) 
\end{equation}
with $\tilde{\eta}$ behaving as $\eta$ and $\tilde{p}_0$ having the dimensions of a pressure divided by length to the power ${\alpha}$. From \eqref{eq:A4} and \eqref{eq:A5}, we see that to have $||\tilde{p}||_{L^1} \underset{\zd \rightarrow 0}{\longrightarrow} 0$, it must be $\alpha > -1$.
Thus, on the galactic plane we have 
\begin{equation}
    \label{eq:A7}
    \pzz(r,0)=\tilde{p}_0(r)\zd^{\alpha-2}\tilde{\eta}''(r, 0)=\frac{\tilde{\eta}''_0(r)}{\zd^2}\tilde{p}(r,0)\,,
\end{equation}
where the prime indicates differentiation w.r.t. $z$. We can now substitute the previous result in \eqref{eq:A1} evaluated on the galactic plane to obtain
    \begin{align}
    \nonumber
    \frac{\textoperatorname{Ł}_{D,r}^2}{2r^2}=& \frac{\textoperatorname{Ł}_{D,r}}{r}(r\vK)_{,r} -\frac{(\vh^2)_{,r}}{r} +\frac{\zd^{\alpha}\dot{\tilde{p}}_0(r) \Sigma'(r)/(2\pi \zd)}{\Sigma(r)^2/(4\pi^2 \zd^2)} -\frac{\zd^{\alpha}\ddot{\tilde{p}}_0(r) +\zd^{\alpha}\dot{\tilde{p}}_0(r)/r +\zd^{\alpha-2}\tilde{\eta}''_0(r)\tilde{p}_0(r)}{\Sigma(r)/(2\pi \zd)} \\
    =&\frac{\textoperatorname{Ł}_{D,r}}{r}(r\vK)_{,r} -\frac{(\vh^2)_{,r}}{r}\nonumber\\ &\hbox to18mm{\hfil}+\frac{1}{\Sigma(r)}\left[\zd^{\alpha+1}\left(2\pi\frac{\Sigma'(r)}{\Sigma(r)}\dot{\tilde{p}}_0(r)-\frac{1}{2}\ddot{\tilde{p}}_0(r) -\frac{\dot{\tilde{p}}_0(r)}{2r} \right) -\frac{\zd^{\alpha-1}}{2\pi}\tilde{\eta}''_0(r)\tilde{p}_0(r)\right] \, ,    \label{eq:A8}
\end{align}
where the dot indicates the differentiation w.r.t.\ $r$ for single-variable functions. We note that the term in $\zd^{\alpha+1}$ necessarily vanishes given the constraint $\alpha>-1$, whilst the term proportional to $\zd^{\alpha-1}$ entails more subtlety. Indeed, for $\alpha>1$, or if $\tilde{\eta}$ is such that $\tilde{\eta}''_0(r)\equiv0$, this term also vanishes, and thus we find exactly \eqref{eq:PracticalLDr}. Moreover, the combination $\tilde{\eta}''_0(r)\neq0, \alpha<1$ is physically {not possible}, since it would give divergent frame-dragging at all radii, (i.e., $\textoperatorname{Ł}_{D,r}(r, 0)$ diverges at each point). Therefore, we are left to consider the last possible combination, namely, $\tilde{\eta}''_0(r)\neq0, \alpha=1$.  To check whether this case is realised we start by noting that a condition on the profile of $\tilde{p}$ around $z=0$ can be obtained from the integrability condition of \eqref{eq:EoMrapp} and \eqref{eq:EoMzapp}. Indeed, we find
\begin{equation}
\label{eq:A9}
    \frac{\pr\rhodz -\pz\rhodr}{\rhod^2}=\frac{(\vh^2)_{,z}}{r}  +\Omega_{,r}\textoperatorname{Ł}_{D,z}-\Omega_{,z}\textoperatorname{Ł}_{D,r}
\end{equation}
We now take $\partial_z$\eqref{eq:A9} on the galactic plane,
so that $\rhodz(r,0) = \pz(r,0) =\Omega_{,z}(r,0) =\textoperatorname{Ł}_{D,z}(r,0)=0$, to get
\begin{equation}
\label{eq:A10}
    \frac{\pr\rho_{d,zz} -\pzz\rho_{d,r}}{\rho_d^2}=\frac{(\vh^2)_{,r}}{r^2} +\Omega_{,r}\left(\frac{\textoperatorname{Ł}_{D,r}}{r}-\textoperatorname{Ł}_{,rr}\right) -\vKzz\frac{\textoperatorname{Ł}_{D,r}}{r}\, ,
\end{equation}  
where we have used the EFE \eqref{eq:tphiexp} and the spherical symmetry of $\vh(R)$, so that $(\vh^2)_{,zz}=(\vh^2)_{,r}/r$ when evaluated on the galactic plane. Furthermore, we know from the Newtonian dynamics that $\vd^2$ is given by \cite{Binney_2008}
\begin{align}
\label{eq:A11}
    v_d^2(r,z)=\frac{G\Md}{2\rd^3}r^2 \left[I_0\left(\frac{r}{2\rd}\right)K_0\left(\frac{r}{2\rd}\right) -I_1\left(\frac{r}{2\rd}\right)K_1\left(\frac{r}{2\rd}\right)\right] -2\pi G\frac{r}{\rd}\Sigma(r)f(r,z;\zd) \, ,
\end{align}
with  $\lim_{\zd \rightarrow 0}f(r,z;\zd) =|z|+O(z^2)$, that is 
$\lim_{\zd \rightarrow 0}f_{,zz}(r,z;\zd) = 2\delta(z)+O(z^0)$.
Therefore, we can write
\begin{equation}
\label{eq:A14}
    f_{,zz}(r,z;\zd)=\frac{2}{\zd}\bar{\eta}(r, z/\zd)\,,
\end{equation}
with some suitable
\begin{equation}
\label{eq:A15}
    \int_{-\infty}^{+\infty}\bar{\eta}(r, s)ds=1\, .
\end{equation}
Moreover, since $\vK^2=\vd^2+\vh^2$ we directly obtain 
\begin{equation}
    \vKzz=\frac{\vd v_{d,zz}+\vh \vhzz}{\vK} \, ,
\end{equation}
on the galactic plane. Thus we can then write
\begin{align}
\vKzz(r,0)&=-\frac{1}{2\vK(r,0)} 2\pi G\frac{r}{r_d}\Sigma(r) \frac{2}{\zd}\bar{\eta}(r,0) +\frac{\vh(r,0)\vhr(r,0)}{r\vK(r,0)}\label{eq:A16} \\ &=\frac{1}{\vK(r,0)}\left[-2\pi G\frac{r\Sigma(r)\bar{\eta}(r,0)}{r_d z_d}+\frac{(\vh^2)_{,r}(r,0)}{2r}\right]\,. \nonumber
\end{align}
Now, we combine the profiles for $\rhod,\, \tilde{p}, \, \text{and} \, \vK$, and substitute in \eqref{eq:A10}, whilst choosing $\eta(r,s):=1/\cosh s$ in order to have a twice differentiable function. We find
\begin{align}
\label{eq:A17}
    -\frac{2\pi \zd^{\alpha-1}}{\Sigma(r)}&\left[\dot{\tilde{p}}_0(r) + \tilde{\eta}''_0(r)\tilde{p}_0(r)\frac{\Sigma'(r)}{\Sigma(r)}\right]\nonumber\\ &=\frac{-[\zd^{\alpha}\dot{\tilde{p}}_0(r)\,\Sigma(r)/(2\pi \zd^3) +\zd^{\alpha-2}\tilde{\eta}''_0(r)\tilde{p}_0(r)\,\Sigma'(r)/(2\pi \zd)]}{\Sigma(r)^2/(4\pi^2 z_d^2)}\nonumber\\
    &= \frac{(\vh^2)_{,r}}{r^2}\; + \;\Omega_{,r}\left(\frac{\textoperatorname{Ł}_{D,r}}{r}-\textoperatorname{Ł}_{,rr}\right) -\frac{1}{\vK}\left(\frac{\vh \vhr}{r} -2\pi G\zd^{-1} \frac{r\Sigma(r)\bar{\eta}_0(r)}{r_d}\right)\frac{\textoperatorname{Ł}_{D,r}}{r}\, .
\end{align}
We note that \eqref{eq:A17} can be satisfied for $\zd\rightarrow0$ only if $\alpha = 0$ and
\allowdisplaybreaks
\begin{align}
    \dot{\tilde{p}}_0(r)+\tilde{\eta}''_0(r)\tilde{p}_0(r)\frac{\Sigma'(r)}{\Sigma(r)} =-G\frac{\Sigma(r)^2\bar{\eta}_0(r)\textoperatorname{Ł}_{D,r}(r,0)}{r_d r\vK(r,0)} \, . \label{p tilde cond}
\end{align} 
Therefore, by substituting $\alpha=0$ in \eqref{eq:A8}, we necessarily conclude that $\tilde{\eta}''_0(r)\equiv0$, for any physical system with nonsingular frame-dragging. This confirms that \eqref{eq:A8} reduces to \eqref{eq:PracticalLDr} for $\zd\rightarrow0$, so that the general relativistic corrections to the pressure play no role in our calculations for a thin disc galaxy.

Finally, interestingly this result is true despite the fact that the pressure in the full GR system differs from the Newtonian case, being given by
\begin{align}
    p(r,z)=p\Z{N}+\tilde{p}=\frac{\pi}{2}G\frac{\Sigma(r)^2}{\cosh{(z/z_d)}} -\frac{G}{r_d}\tilde{\eta}(r, z/z_d)\int\frac{dr}{r}\bar{\eta}_0(r)\frac{\Sigma(r)^2\textoperatorname{Ł}_{D,r}(r,0)}{\vK(r,0)}\, . 
\end{align}
However, such corrections do not affect our results as we have proved that they play no role for an infinitely thin disc -- a common assumption in galaxy models: see, e.g., \cite{Persic_1996} and discussion therein.

\end{document}